\documentstyle[amsmath,amssymb,epsfig,12pt]{article}

\begin{document}
\setlength{\parskip}{0.45cm}
\setlength{\baselineskip}{0.75cm}
%
%
%
\begin{titlepage}
\setlength{\parskip}{0.25cm}
\setlength{\baselineskip}{0.25cm}
\begin{flushright}
DO-TH 07/04\\
\vspace{0.2cm}
May 2007
\end{flushright}
\vspace{1.0cm}
\begin{center}
\Large
{\bf The Lamb shift contribution of very light millicharged particles}
\vspace{1.5cm}

\large
M.~Gl\"uck, S.~Rakshit, E.\ Reya\\
\vspace{1.0cm}

\normalsize
{\it Universit\"{a}t Dortmund, Institut f\"{u}r Physik}\\
{\it D-44221 Dortmund, Germany}

\vspace{1.5cm}
\end{center}

\begin{abstract}
\noindent The leading order vacuum polarization contribution of 
very light millicharged fermions and scalar (spin--0) particles 
with charge $\varepsilon e$
and mass $\mu$ to the Lamb shift of the hydrogen atom is shown to
imply {\em universal}, i.e.~$\mu$--independent, upper bounds on
$\varepsilon:\varepsilon$ 
\raisebox{-0.1cm}{$\stackrel{<}{\sim}$} $10^{-4}$ for 
$\mu$ \raisebox{-0.1cm}{$\stackrel{<}{\sim}$} 1 keV in the case
of fermions, and for scalars this bound is increased by a factor
of 2.  This is
in contrast to expectations based on the commonly used approximation
to the Uehling potential relevant only for conventionally large
fermion (and scalar) masses.
\end{abstract}
\end{titlepage}


The recent observation \cite{ref1} of an optical rotation of linearly
polarized laser light generated in vacuum by a magnetic field may be due
\cite{ref2} to the photon initiated pair production of very light charged
fermions with mass $\mu\simeq 0.1$ eV and charge $\varepsilon e$
where $\varepsilon\simeq 10^{-6}$.  There exist, however, very strong
astrophysical, cosmological and laboratory constraints 
\cite{ref3, ref4, ref5} 
which exclude 
the quoted values of $\varepsilon$ and $\mu$.  Some of these constraints
may nevertheless be relaxed in specific paraphoton scenarios \cite{ref5}.
Further constraints on the mass and charge of light charged fermions may
be obtained from their leading order vacuum polarization contribution 
\cite{ref3,ref7,ref8} to the Lamb shift of the hydrogen atom
\begin{equation}
\delta E = E(2 S_{1/2}) - E(2P_{1/2})
\end{equation}
or from their higher order contribution \cite{ref7} to the anomalous
magnetic moment of the muon.  The Lamb shift constraints for
conventionally large fermion masses studied in
\cite{ref3,ref7,ref8} were, as ususal,  based on the commonly used
{\em approximate} leading order vacuum polarization contribution
\begin{equation}
\delta E_{\rm VP} \simeq -\frac{\alpha^5 m_e}{30\pi} \, 
 \left( \frac{m_e}{\mu}\right)^2 \varepsilon^2 \, .
\end{equation}
However, the upper bound presented in \cite{ref3,ref9} is not
appropriate for
$\mu$ \raisebox{-0.1cm}{$\stackrel{<}{\sim}$} 1 keV.
Even much lower values of $\mu$, $\, \mu<1$ eV, were  considered in \cite{ref2},
for example, 
and it is therefore necessary to study the consequences of going beyond
this standard approximation relevant only for $\mu >\alpha m_e\,.$

The lowest order Coulomb interaction $V=-\alpha/r$ for a point nucleus
is modified at the 1-loop level according to $V=-\alpha/r+\delta V$ where
\cite{ref10,ref11}
\begin{equation}
\delta V(r) = -\frac{\alpha}{r}\,\, \frac{2\alpha\varepsilon^2}{3\pi}
 \int^{\infty}_1 du\, e^{-2\mu ru}
  \left( 1+\frac{1}{2u^2} \right)\frac{\sqrt{u^2-1}}{u^2}
\end{equation}
and $\alpha=e^2/4\pi = 1/137.036$.  This leading order Uehling potential
yields, instead of the approximate equation (2), the {\em exact} leading
order expression
\begin{equation}
\delta E_{\rm VP} = \int_0^{\infty} dr\, r^2\delta V(r) 
  \left[R_{20}^2(r) - R_{21}^2(r)\right]
\end{equation}
with the normalized radial hydrogen wave functions $R_{n\ell}$ given by
\begin{equation}
R_{20}(r) = \frac{1}{\sqrt{2}}\,\, \frac{1}{a^{3/2}}
   \left( 1-\frac{\rho}{2}\right) e^{-\rho/2}\,, \quad
R_{21}(r) = \frac{1}{2\sqrt{6}}\,\,\frac{1}{a^{3/2}}\rho\, e^{-\rho/2}
\end{equation}
where $\rho= r/a$ and $a^{-1}=\alpha m_e$.  The integration over $r$ in (4)
yields 
\begin{equation}
\delta E_{\rm VP} = -\frac{4\alpha^3m_e}{3\pi}\, \varepsilon^2\, 
  \alpha^{*2} I(\alpha^*)
\end{equation}
where
\begin{equation}
I(\alpha^*) = \int_1^{\infty} du\left(1+\frac{1}{2u^2}\right)\,\,
  \frac{\sqrt{u^2-1}}{(\alpha^*+2u)^4}
\end{equation}
with the effective coupling $\alpha^*=\alpha m_e/\mu$.  Requiring this leading
order vacuum polarization contribution $|\delta E_{\rm VP}|/2\pi\hbar$ to
the Lamb shift not to exceed 0.01 MHz, corresponding to a 1$\sigma$ discrepancy
between the measured and calculated shifts \cite{ref12,ref13}, yields the 
constraints on $\varepsilon$ shown in Fig.~1.  
(Imposing a $2\sigma$ error, i.e.~increasing the experimental uncertainty
to 0.02 MHz, increases the upper bound on $\varepsilon$ by a factor
of $\sqrt{2}\, $.)
It is interesting to note that,
contrary to what one might naively expect, it turns out that the integral
$I(\alpha^*)$ in (6) entirely compensates the large enhancement factor 
$\alpha^{*2}$ already at
$\mu$ \raisebox{-0.1cm}{$\stackrel{<}{\sim}$} 1 keV due 
$\alpha^{*2}I(\alpha^*)=\frac{1}{24}$ for $\alpha^*\gg 1$.
(This asymptotic result is easily obtained by observing that the maximum
of the integrand in (7) moves to $\infty$ as $\alpha^*\to\infty$ where the
integrand reduces to $u(\alpha^*+2u)^{-4}$.)  Therefore the upper bound
in Fig.~1 saturates below $\mu\simeq 1$ keV which implies a rather moderate
but {\em universal}, i.e.\ $\mu$--independent, upper bound
\begin{equation}
\varepsilon \leq 1.085 \times 10^{-4}\quad\quad{\rm for}\quad\quad
  \mu \lesssim 1 \, {\rm keV}\, .
\end{equation}
Although in the different context of electronic vacuum polarization corrections
to the energy levels of exotic atoms, the exact result (7) and its
asymptotic limit was previously presented \cite{ref14} in a different 
form.\footnote{The integral in (7) as well as the ones for the 
(physically irrelevant) individual 2$S$ and 2$P$ contributions can in 
general be calculated analytically.  The rather lengthy analytic
expressions of the latter ones, together with their asymptotic
limits ($\alpha^*\gg 1$), are given in Appendix A of \cite{ref14}
by identifying $\kappa_2=\alpha^*/2$.  Although these expressions
are less transparent for understanding our universal (scaling)
upper bound on $\varepsilon$, it should be mentioned that the 
individual 2$S$ and 2$P$ contributions do not scale asymptotically
but increase as $\ln \alpha^*-\frac{7}{3}$ and 
$\ln\alpha^*-\frac{8}{3}$, respectively, in units of 
$-\alpha^3\varepsilon^2 m_e/6\pi$.  The advantage of our more direct
approach is that it can be immediately applied also to the case of
scalar particles studied below.}
Despite the fact that our eqs.~(6) and (7) are particularly suited for
{\em directly} understanding the reasons for the universal bound on
$\varepsilon$ discussed above, we also present for completeness the
analytic result for the integral in (7):
\begin{eqnarray}
\alpha^{*2}\, I(\alpha^*) & = & \frac{1}{8\kappa^3}
 \Bigg[2\pi+\frac{\kappa}{3(\kappa^2-1)^2}
  \Big( -12 + 22\, \kappa^2-\frac{13}{2}\, \kappa^4+\kappa^6\Big)\nonumber
\\
& & -\frac{1}{(\kappa^2-1)^2}
   \Big(4-10\ \kappa^2+\frac{15}{2}\, \kappa^4\Big)\, L\Bigg]
\end{eqnarray}
with $\kappa=\alpha^*/2$ and 
$L =\frac{\ln(\kappa +\sqrt{\kappa^2-1})}{\sqrt{\kappa^2-1}}$
relevant for $\kappa > 1$, and 
$L = \frac{{\rm arc}\,\cos \kappa}{\sqrt{1-\kappa^2}}$ for $\kappa<1$.
This analytic result has originally been given explicitly in 
\cite{ref15} in the context of mesonic atoms.  For a comprehensive 
review on radiative QED corrections to hydrogenlike atoms we refer
the reader to \cite{ref16}.
 
The dashed curve in Fig.~1 displays the upper bound as obtained from using
\cite{ref3,ref8} the approximate expression (2) which, imposing 
$|\delta E_{\rm VP}|/2\pi\hbar \leq 0.01$ MHz, implies
$\varepsilon \leq \mu/$ \mbox{(26.62 MeV)} as was assumed to hold for 
$\mu > 1$ keV \cite{ref8}.  It can be already seen from Fig.~1 that this
approximate bound is not correct for $\mu$ much below $10^5$ eV.
To illustrate this more clearly, we confront in Fig.~2 the approximate 
and exact upper bounds on $\varepsilon$, as obtained from (2) and (6),
respectively, using a linear scale for $\varepsilon$.  Their convergence
at 
$\mu$ \raisebox{-0.1cm}{$\stackrel{>}{\sim}$} $10^5$ eV is due to
$I(\alpha^*)\to\frac{1}{40}$ for $\alpha^*\to 0$ in (7). 

It should be mentioned that corrections to the anomalous magnetic moments
of electrons and muons, or to the hyperfine splitting, for example, are
genuinely {\em{sub}}leading and thus the resulting bounds are much
weaker \cite{ref7}.  Furthermore, considerations of the level shifts of
exotic atoms \cite{ref14} obviously can not improve the upper bounds on
$\varepsilon$ as obtained for the hydrogen atom, since the associated
relative theoretical uncertainties are generally more significant
\cite{ref17}.

Next we also calculate the exact upper bound for light millicharged
scalar \mbox{(spin--0)} particles of mass $\mu$. In this case the standard
leading order fermionic QED contribution 
$x(1-x)\ln \left[ 1-x(1-x)\, q^2/\mu^2\right]$
appearing in the integrand of the Feynman--parameter integral in the
vacuum polarization tensor \cite{ref10,ref18} has to be replaced by
$\frac{1}{8}(2x-1)^2\ln\left[1-x(1-x)\, q^2/\mu^2\right]$.
Performing now the exact integrations as outlined in \cite{ref10}, for
example, one arrives at 
\begin{equation}
\delta V_s(r) = -\frac{\alpha}{r}\,\, 
  \frac{\alpha \varepsilon^2}{6\pi} \int_1^{\infty}du\, 
    e^{-2\mu ru}\left(1-\frac{1}{u^2}\right)\, 
     \frac{\sqrt{u^2-1}}{u^2}
\end{equation}
for the 1--loop correction of the leading order Coulomb interaction
$V=-\alpha/r +\delta V_s$.  The calculation of this vacuum polarization
contribution to the Lamb shift is now analogous to the above case of
fermions, and the {\em exact} leading order contribution of scalars
reads
\begin{equation}
\delta E_{\rm VP}^s = -\frac{\alpha^3 m_e}{3\pi}\, \varepsilon^2\,
  \alpha^{*2} I_s(\alpha^*)
\end{equation}
where
\begin{equation}
I_s(\alpha^*) = \int_1^{\infty} du\left(1-\frac{1}{u^2}\right)\,\,
  \frac{\sqrt{u^2-1}}{(\alpha^*+2u)^4}\, \, ,
\end{equation}
to be compared with the fermionic result in (6) and (7).  Imposing
the same experimental upper bound on 
$|\delta E_{\rm VP}^s|/2\pi\hbar$ of 0.01 MHz as above for fermions
yields the constraints on $\varepsilon$ shown in Fig.~1 for scalars.
The scalar upper bound in Fig.~1 again saturates below $\mu\simeq 1$
keV due to 
$\alpha^{*2}I_s(\alpha^*) = \frac{1}{24}$ for
$\alpha^*\gg 1$, which is the same asymptotic result as in the fermionic
case for $\alpha^{*2}I(\alpha^*)$.  Therefore, since the scalar
contribution in (11) is 4 times smaller than the fermionic one in
(6), the {\em universal}, i.e. $\mu$--independent, upper bound for
the charge of scalar particles is a factor of $\sqrt{4}$ larger than
the one given in (8), i.e.
\begin{equation}
\varepsilon < 2.17 \times 10^{-4}\quad {\rm for}\quad 
  \mu \lesssim 1\,{\rm keV}\, .
\end{equation}
The dashed curve in Fig.~1 for scalars displays the upper bound as
obtained from utilizing the approximate expression \cite{ref7}
\begin{equation}
\delta E_{\rm VP}^s \simeq -\frac{\alpha^5 m_e}{240\pi}\,\,
   \left(\frac{m_e}{\mu}\right)^2\varepsilon^2
\end{equation}
which is obtained from (11) using $I_s(\alpha^*)\to \frac{1}{80}$
for $\alpha^*\to 0$.  (Notice that this result is a factor of 8 smaller
than the fermionic approximate result in (2).)  It can be seen from
Fig.~1 that the approximate scalar bound is not correct for $\mu$ 
much below $10^5$ eV as in the fermionic case. Due to the different
integrand in (12), as compared to (7), the difference between the 
approximate and exact upper bounds is more pronounced than for fermions
which is illustrated more clearly in Fig.~3 (to be compared with 
Fig.~2).
Finally it should be mentioned that the scalar scenario is 
experimentally somewhat favored (cf.~Table V of \cite{ref2}) over
the fermionic one.

Again, as in the fermionic case, our eqs.~(11) and (12) are particularly
suited for {\em directly} understanding the reasons for our universal 
bound (13) on $\varepsilon$.  Nevertheless we also present for completeness
the analytic result for $I_s$ in (12) as obtained from a straightforward
integration:
\begin{eqnarray}
\alpha^{*2}\, I_s(\alpha^*) & = & \frac{1}{8\kappa^3}
 \Bigg[-4\pi+\frac{\kappa}{\kappa^2-1}
  \Big( -8 + \frac{20}{3}\, \kappa^2+ \frac{1}{3}\, \kappa^4\Big)\nonumber
\\
& & -\frac{1}{2(\kappa^2-1)}
   \Big(16-24\ \kappa^2+ 6 \kappa^4\Big)\, L\Bigg]
\end{eqnarray}
with $\kappa$ and $L$ as in (9).

To conclude, we have shown that the contribution of very light fermions
with charge $\varepsilon e$ and mass $\mu$ to the Lamb shift of the 
hydrogen atom implies a {\em universal}, i.e.\ $\mu$--independent, upper
bound 
$\varepsilon$ \raisebox{-0.1cm}{$\stackrel{<}{\sim}$} $10^{-4}$ for
$\mu$  \raisebox{-0.1cm}{$\stackrel{<}{\sim}$} 1 keV.  This result is
only obtainable by utilizing the exact expression for the Uehling potential
rather than its standard approximation \cite{ref11,ref18}
commonly used in QED.  Since the scalar spin--0 scenario is experimentally 
somewhat favored \cite{ref2}, we calculated the exact upper bound for
millicharged light scalar particles as well, which turns out to be a
factor of 2 larger for 
$\mu$ \raisebox{-0.1cm}{$\stackrel{<}{\sim}$} 1 keV than for fermions.
\vspace{1.0cm}

\noindent{\underline{\bf Acknowledgements}}

\noindent This work has been supported in part
by the `Bundesministerium f\"ur Bildung und Forschung', Berlin/Bonn.

\clearpage
\begin{figure}
\epsfig{file=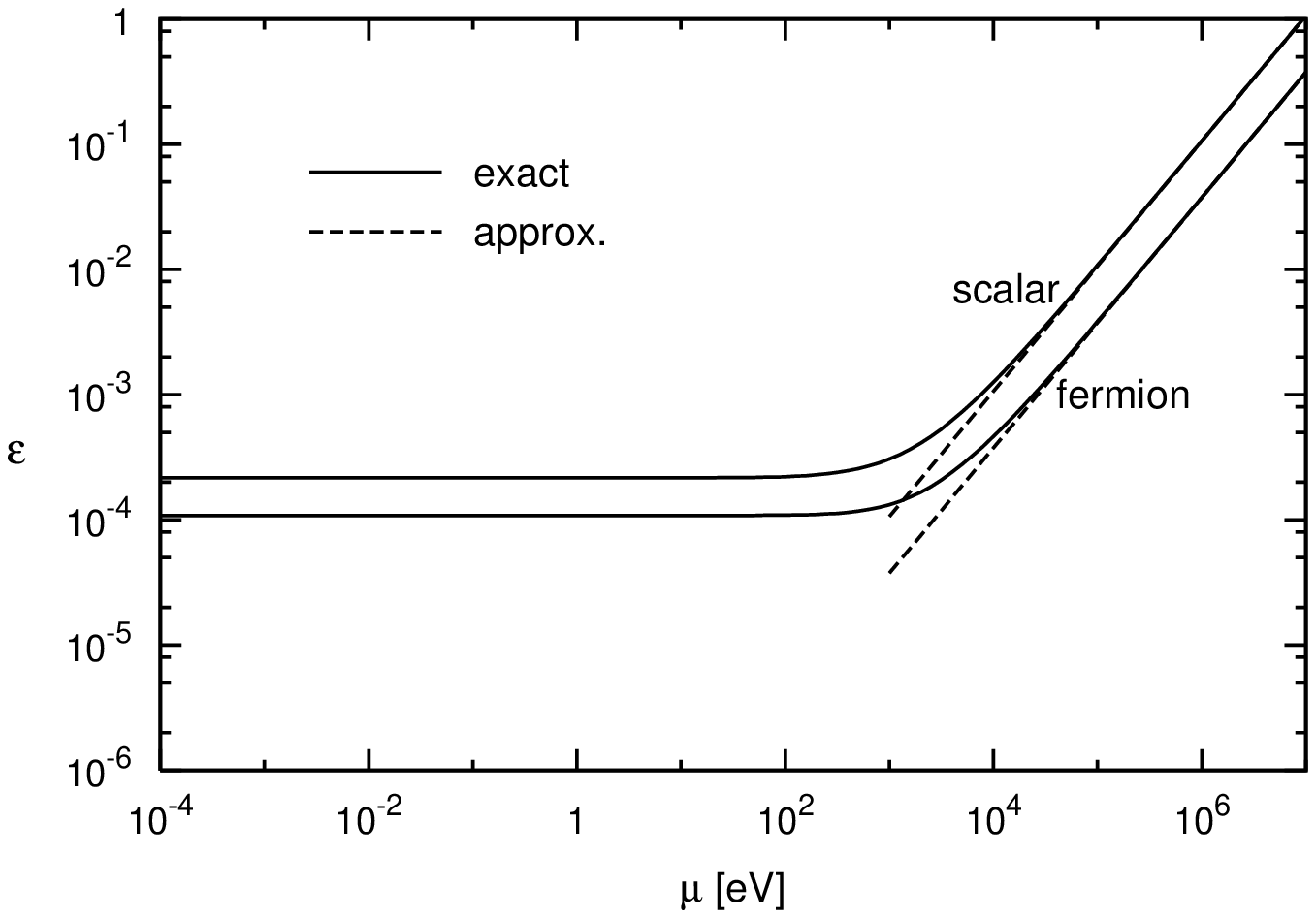, width=12cm}
\caption{Allowed upper bounds of $\varepsilon$ according to the exact
leading order contribution (6) and (11) of fermions and scalars,
respectively, to the Lamb shift of the hydrogen atom
corresponding to a $1\sigma$ discrepancy of 0.01 MHz between theory and
experiment.  The dashed `fermion' curve corresponds to the bound suggested in
\cite{ref3,ref8} being based on the approximate expression (2) as
discussed in the text. The dashed `scalar' curve is obtained from 
the approximate expression (14). (In order to avoid any confusion it 
should be stressed that the area above the respective curves is excluded.)}
\end{figure}

\clearpage
\begin{figure}
\epsfig{file=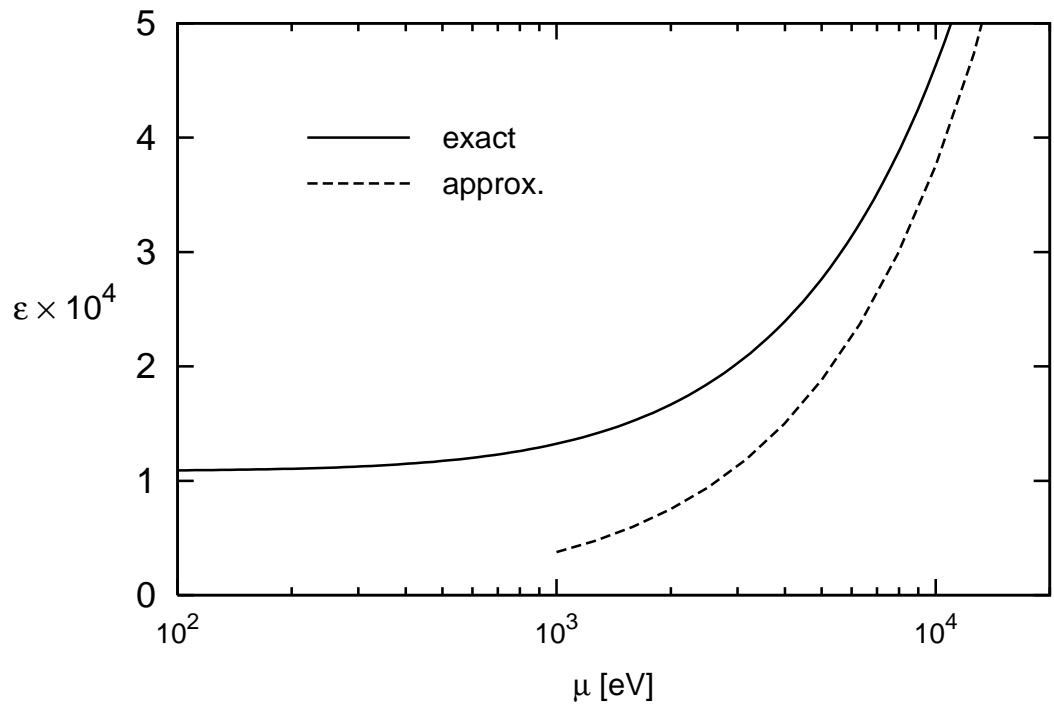, width=12cm}
\caption{As in Fig.~1 for fermions but using a linear scale for 
$\varepsilon$.}
\end{figure}

\clearpage
\begin{figure}
\epsfig{file=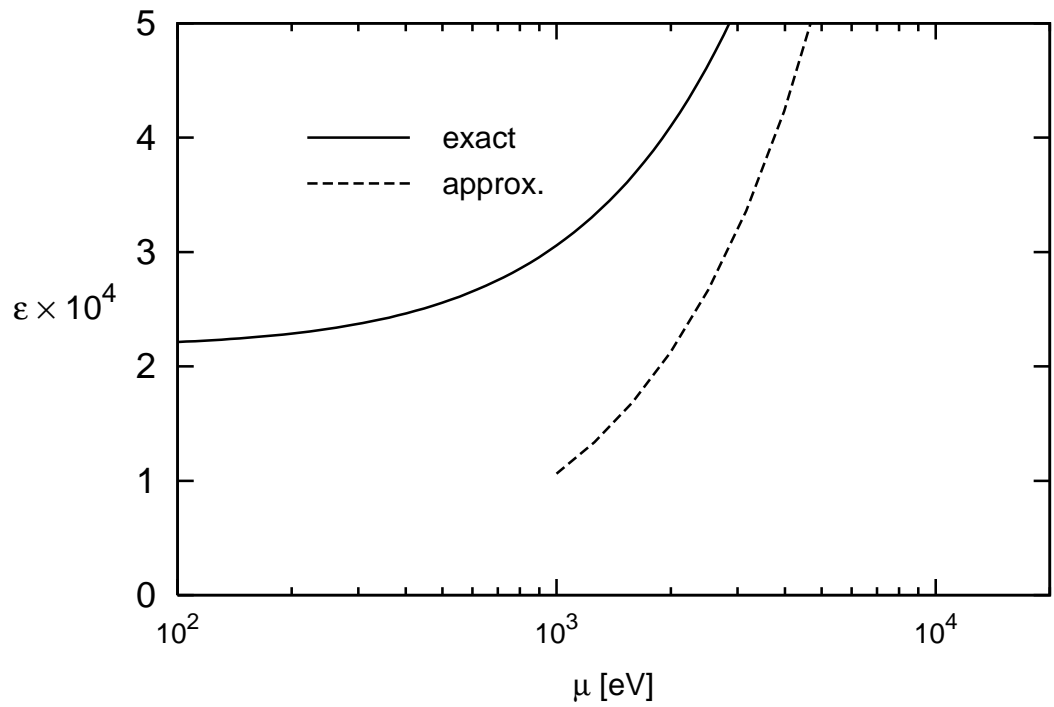, width=12cm}
\caption{As in Fig.~1 for scalars but using a linear scale for
$\varepsilon$.}
\end{figure}

\end{document}